\documentclass[prd, preprint, 11pt]{revtex4-1}

\usepackage{amsmath}
\usepackage{amssymb}
\usepackage{setspace}
\usepackage{graphicx}
\usepackage{natbib}
\usepackage{float}

\begin{document}
 
 %

\begin{center}
 { \large {\bf Information and the Foundations of Quantum      Theory}}


\vskip 0.2 in

{\large{\bf Angelo Bassi$^a$, Saikat Ghosh$^b$ and Tejinder  Singh$^c$}}

\medskip
{\it $^a$Department of Physics, University of Trieste, Strada Costiera 11, 34151 Trieste, Italy}\\
{\it $^b$Department of Physics, Indian Institute of Technology Kanpur,  Kanpur 208016, India }\\
{\it $^c$Tata Institute of Fundamental Research,}
{\it Homi Bhabha Road, Mumbai 400005, India}\\
\medskip

\end{center}

\centerline{\bf ABSTRACT}

\smallskip

\setstretch{1.1}

\noindent We believe that the hypothesis `it from bit' originates from the assumption that probabilities have a fundamental, irremovable status in quantum theory. We argue against this assumption and highlight four well-known reformulations / modifications of the theory in which probabilities and the measuring apparatus do not play a fundamental role. These are: Bohmian Mechanics, Dynamical Collapse Models, Trace Dynamics, and Quantum Theory without Classical Time. Here the `it' is primary and the `bit' is derived from the `it'. 

\medskip

{\noindent {\bf \it This essay received the second prize in the FQXi 2013 Essay Contest `It from Bit, or Bit from It?'}}

\smallskip


\section{Introduction}
\noindent In the standard approach to quantum theory, the state of the quantum system is described by the wave function, whose evolution is given by the deterministic Schr\"odinger equation. However, when a measurement is performed on this system by a classical apparatus, the outcome is not deterministically related to the initial state. Instead, in any specific realization of the measurement, one or the other outcome occurs with a certain probability. 

This sudden onset of probabilities in a system which is otherwise evolving deterministically while it is `unmeasured', has sometimes been used to accord probabilities a fundamental, {\bf irremovable} status in quantum theory. It has been suggested that the system or the object being measured upon (say an electron) does not have definite properties before the measurement, but rather resides in some probabilistic realm, and acquires well-defined physical properties only upon measurement (an act of seeking information). This outlook, namely that {\bf the definitive properties of a quantum system somehow become a reality only when information about them is sought by an act of measurement}, is perhaps the foundation of the hypothesis `it from bit'. 

If we dwell on the above reasoning, it does not take much effort to narrow down to two places where the argument is weak enough to be essentially flawed:

$\bullet$ One is the so-called classical measuring apparatus, and the other is the status of probabilities. When is an apparatus classical? Strictly speaking, we do not quite know. Quantum theory does not say how large an object must be, before it can be said to obey the rules of Newtonian mechanics. Should its mass be a billion a.m.u. or a trillion a.m.u. or something else? The theory does not tell us. And why should the theory have to depend on its own limit [the classical apparatus, whatever that might mean] in order to complete the description of the formalism?  Indeed, when something so well-defined mathematically such as the Schr\"{o}dinger equation and the commutation relations are supplemented by something as vague as a `measuring apparatus', we should smell rat, and know that things are amiss! We should look for a first principles holistic description of quantum theory which does not make explicit reference to a {\it classical} measuring apparatus. 

$\bullet$ And secondly, there is no place for probabilities in a system evolving deterministically, and for which the initial conditions [the wave function] are exactly known. Probabilities arise when there is a pre-given sample to choose from, and the initial state is not precisely known. Such of course is the case in statistical mechanics, and in coin tossing. But not so in quantum theory - the only known physical theory where probabilities come into play without there being a sample of initial states. 

The only way to overcome this illogical state of affairs is to look for a more complete formulation of the theory which offers a mathematcal explanation for the random outcomes of measurements, thus getting rid of the fundamental inexplicable status of probability. 

This could be achieved if the evolution is deterministic but the initial state is not precisely known [{\bf Bohmian mechanics}]. Or it could happen if there is a stochastic nonlinear aspect to the evolution, over and above the Schr\"{o}dinger evolution, which becomes significant during measurement, and dynamically causes collapse of the wave function into one or the other outcomes, in accordance with the Born probability rule [{\bf dynamical collapse models}]. Remarkably enough, Bohmian mechanics as well as collapse models also do away with any direct reference to the classical measuring apparatus. A measurement is nothing but another aspect of dynamical evolution.

Figure 1 makes it abundantly clear that one should not attribute the emergence of random outcomes from a deterministic evolution to some mysterious and mathematically ill-defined `measurement'. The evolution should either be supplemented by random initial conditions [{\bf B.} Bohmian Mechanics] or by a stochastic  aspect to the evolution itself [{\bf S.} Schr\"{o}dinger + Stochastic Evolution]. Such restoration of mathematical completeness is how physicists generally go about amending shortcomings in a physical theory.

\begin{figure} [ht]
 {\includegraphics[width=\textwidth]{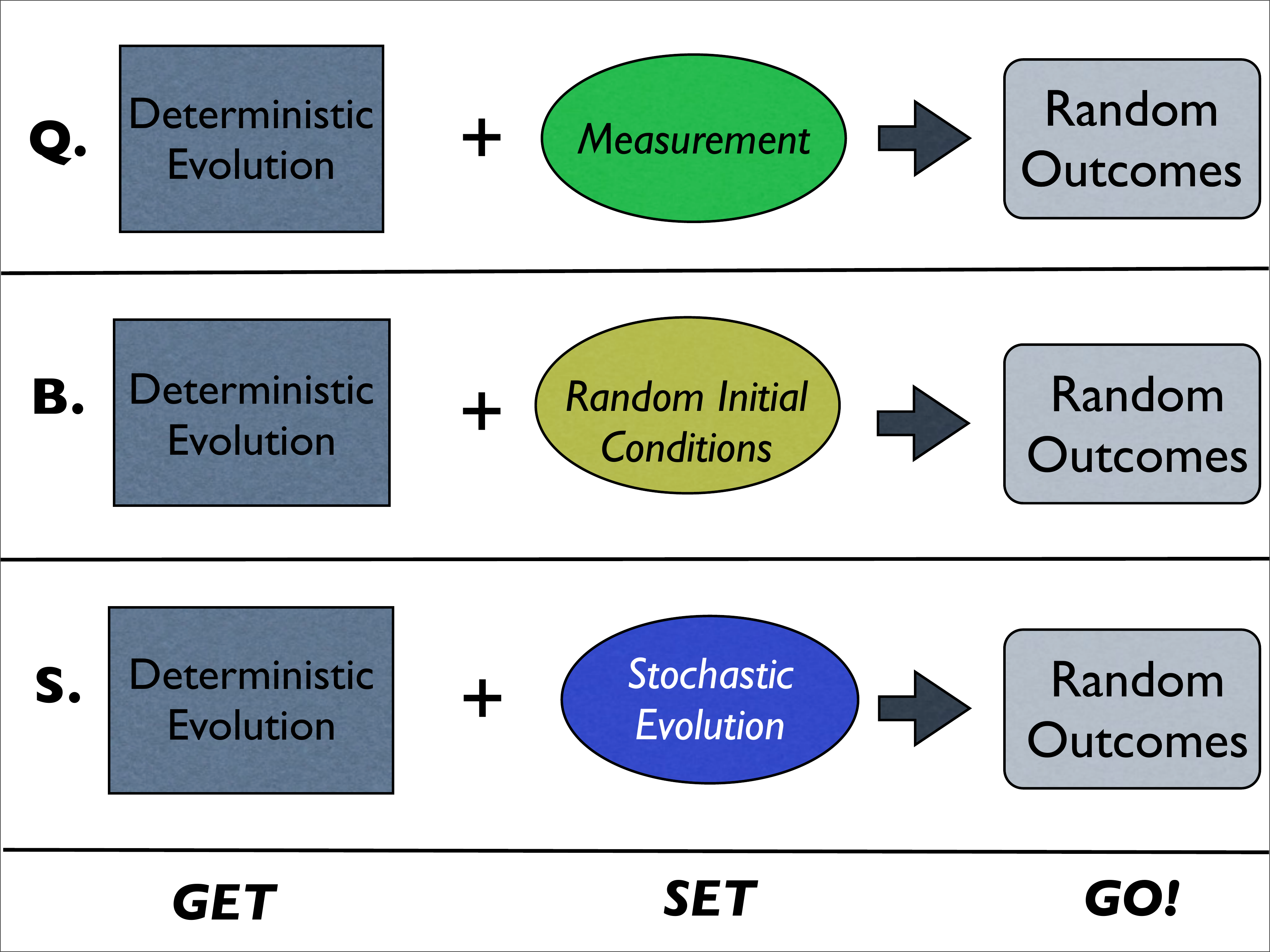}}
\caption{The quantum  {\bf Q.} gets a makeover: {\bf B.} Bohmian Mechanics {\bf S.} Stochastic + Deterministic Evolution}  
\end{figure}%

 When such a new formulation or a modified theory is presented, the `it from bit' vanishes into thin air, literally! No longer does the existence or reality of physical systems depend on measurements / information/ the questions we ask about them. The `it' necessarily comes first. Below, we recall four diffent routes towards a more complete formulation of quantum theory, where neither the classical apparatus nor probabilities play a fundamental role, and in each of the routes, the `it' regains its primary status. 

It is sometimes said that it is not necessary to consider reformulations / modifications of quantum theory, because the theory agrees with every experiment performed to date, to test it. While this agreement of theory and experiment is undoubtedly remarkable, one should not forget that there are regions of the parameter space (quantified by number of degrees of freedom in the object being studied) where the theory has not yet been subjected to laboratory tests. Such ongoing / planned tests hold the potential for revealing the need to modify the theory.

\section{Bohmian Mechanics}

First discovered by de Broglie in 1927, rediscovered by Bohm in 1952, and subsequently championed by John Bell, Bohmian Mechanics [BM] is perhaps the most misunderstood reformulation of quantum theory! It is seldom appreciated that BM is perhaps the most compact and elegant way of doing away with the shortcomings of the standard formulation, without changing the theory \cite{Bohm:52, Bohm2:52, Bub:1997, Duerr, Holland}

Put simply, BM is a theory of particles and their trajectories. The particle {\bf IS} the `it' and all information is about the particle. But BM is by no means a return to classical physics and determinism. In this theory, a system of $N$ non-relativistic particles is described by a wave function which lives on configuration space, {\it and} by the {\bf actual} positions of the particles. The positions evolve according to a `guiding equation' which depends on the wave function, while the wave function itself obeys the Schr\"{o}dinger equation.

The initial conditions which determine the trajectories are random, and distributed according to the Born probability rule expressed through the wave function [the so-called quantum equilibrium hypothesis]. But this randomness of the initial conditions in no way takes away the reality of the particles and the trajectories. The `itness' of the particles stands firm, the bit follows from it, and the wave function is secondary. Usefully though, the randomness helps understand the collapse of the wave function, without assigning a central role to the measuring apparatus.

BM explains everything that conventional non-relativistic QM does. But then people ask: what good is a reformulation which gives the same results as the standard theory? [A renowned physicist has gone as far as to call BM `verbal window dressing'.] Such a question would perhaps not be asked if BM had been discovered before the standard theory replete with the Copenhagen interpretation. Instead we might be asking ourselves: what good is the Copenhagen interpretation and the accompanying ad hoc probabilities when we already have BM?!

It is undeniably the case that the chronological order in which different formulations were discovered, and their accompanying sociological impact, has had a great deal to do with the bit preceding the it, in some quarters.

\section{Dynamical collapse of the wave function}

Or it could be that the fundamental evolution is nonlinear and stochastic, which includes the collapse of the wave function together with the standard quantum properties. 
How could we be so certain that in quantum theory, there is only a linear, deterministic, part to the evolution, described by the Schr\"{o}dinger equation?  The history of science has shown that linear theories often are approximations to more fundamental nonlinear theories, like in the case of Newtonian gravitation and general relativity. Therefore it seems natural to seek out nonlinear extensions of the Schr\"{o}dinger equation. 

There is nothing in today's experiments which rules out the inclusion of a stochastic nonlinear aspect in the quantum evolution. All that is required is that such an aspect should be extremely tiny and negligible for microscopic systems. On the other hand the stochastic aspect can well become significant for macroscopic systems. This is not ruled out by experiments; on the contrary, the random nature of outcomes in a measurement suggests such a feature!

This is because stochastic nonlinearity breaks quantum linear superposition during a measurement. When a quantum system interacts with a measuring apparatus, together they behave like a macroscopic system, for which the stochastic nonlinear evolution dominates the linear Schr\"{o}dinger evolution, resulting in random outcomes which do not obey linear superposition, as observed.  

In such a scenario, the `it' is once again primary, being an objective reality described by the quantum state of the system. It is no problem that the evolution of this state is described by a stochastic Schr\"{o}dinger equation. Once one has the collapse in the dynamics, the wave function provides a satisfactory description of physical reality.

The derived `bit' is constituted by all the information we have about this quantum system - conventionally such information is collected through the interaction of this system with a macroscopic object obeying the laws of classical mechanics. It is no surprise if the information we collect is based on outcomes which are probabilistic, for this is an inevitable consequence of the interaction of a highly deterministic [quantum] system with a highly stochastic [classical] system.

Such a modified quantum theory, which combines deterministic evolution with stochastic evolution, has been successfully developed by a group of physicists, since the eighties \cite{Pearle:76, Diosi:89, Ghirardi:86, Gisin:81}. In its currently most advanced version it is known as `Continuous Spontaneous Localization' [CSL] \cite{Ghirardi2:90, Bassi:03, RMP:2012}. This is a stochastic nonlinear Schr\"{o}dinger equation, in which the standard linear evolution is supplemented by a nonlinear stochastic part, described by a Weiner process. The stochastic part enforces significant consequences on the Schr\"{o}dinger evolution. It causes position localization by opposing the quantum spread of the wave function. Such localization is shown to be unimportant for micro-systems, thus explaining their wavy nature. On the other hand the localization is very significant for macro-systems, thus explaining their classical Newtonian behaviour. There is hence a universal dynamics, which on the one hand can explain the quantum nature of atomic and molecular phenomena, and on the other hand explain the classical nature of large objects. And none of this makes any reference to measurement.  

Nor does this universal dynamics leave any place for fundamental, irremovable probabilities. The Born probability rule is shown to be a mathematical consequence of the CSL equation. When a quantum system, which is in a superposition of various eigenstates of the observable being measured, interacts with a classical measuring apparatus, the CSL equation shows that superposition is broken and one or the other outcomes is realized in accordance with this probability rule. This is random determinism: randomness is a fundamental feature of the law of evolution; it does not in any way take away the primary importance of the `it'.

Today, technology is at a stage where CSL is being put to experimental tests in the laboratory, for in the mesoscopic and macroscopic range its predictions  markedly differ from those of quantum theory \cite{RMP:2012}. If CSL is confirmed, then `it' will reign supreme. If CSL is ruled out, Bohmian mechanics will gain centerstage, and `it' will still hold ground!

Of course, having introduced a stochastic element into natural laws, the onus is on CSL to explain where this stochasticity comes from. Else, the criticism that it was invented for the sole purpose of explaining measurements and removing probabilities would be well-founded! 

Perhaps there is a universal stochastic field in nature, of cosmological and / or gravitational origin. This field interacts with all material objects in the manner described by CSL. In fact the universal nature of fluctuations of the gravitational field, and their consequent impact on quantum evolution, has been emphatically highlighted. And this idea is also being put to test in the laboratory.

Alternatively - and this makes the case for `bit from it' ever stronger - the stochastic element is a consequence of coarse graining of a fundamental deterministic theory which describes the evolution of the state of the `it'. Such a theory, from which CSL originates, and to which quantum theory is an approximation, has indeed been developed by Stephen Adler and colaborators, and we briefly allude to it below.

\section{Trace Dynamics}

Why should there be any such thing as `quantization'? Why do we have to be first given a classical theory, and we then use a recipe to obtain the quantum theory by `quantizing' the classical theory? If we have to use a theory's own limit to deduce the theory, this is not the most satisfactory state of affairs, and this is one of the problems Adler's well thought out theory of Trace Dynamics [TD] sets out to address \cite{Adler:94, Adler:04, Adler:06a, Adler-Millard:1996}. TD is a theory of the classical dynamics of Grassmannian matrices. A matrix degree of freedom at a given point in space may be thought of as representing  a particle, and its time evolution, given by Newtonian dynamics, defines a spacetime `trajectory' for the matrix. This matrix is the ultimate `it', and if there were any misgivings caused by the stochasticity explicitly introduced by hand in CSL, those misgivings are removed here. For there is no fundamental stochasticity here, to begin with. 

Why does one start with matrices? Because these matrices, which do not commute with each other, and have arbitrary commutation relations, serve as precursors of the position and momentum operators of quantum theory. One of the most remarkable and unique features of this matrix theory is that it possesses a conserved charge, made out of the sum of the commutators of the matrices and their corresponding momenta. Even though each of these commutators is arbitrary and time-dependent, their sum is conserved! This charge, which has the dimensions of action,  plays a central role in the deduction of quantum theory from TD.

Next, one posits that at the scale at which we perform our laboratory experiments, we do not probe these matrices. Much in the same way in which while studying the macroscopic properties of a gas, we do not probe individual atoms, but only the thermodynamic properties of the coarse-grained system. In the same spirit one constructs the statistical thermodynamics of the dynamical theory of matrices: the matrices are the atoms, and their statistical averaging is the macroscopic gas.

Following the well laid-out laws of equilibrium statistical mechanics, one constructs a probability density distribution, whose equilibrium configuration is derived by maximizing the Shannon entropy. Given this probability distribution, and the invariance of thermodynamic averages under translations in phase space, one is led to an analog of the equipartition theorem. From here emerge the canonical commutation relations of standard quantum theory, satisfied by thermal averages of the underlying matrices (operators). There also emerge the Heisenberg equations of motion satisfied by the thermally averaged position and momentum operators. As in standard quantum theory, there is a Schr\"{o}dinger evolution equivalent to the Heisenberg evolution.  {\bf Quantum theory is the statistical thermodynamics of a classical matrix dynamics.}

Next comes the icing on the cake: we find the `bit' emerging from the `it'. Where there is equilibrium thermodynamics, there are statistical fluctuations [Brownian motion]. While quantum theory corresponds to the equilibrium thermodynamics of the averaged matrix theory, the inclusion of fluctuations results in a modified nonlinear stochastic Schr\"{o}dinger equation of the CSL type! Probabilities are thus the consequence of a stochastic element which has emerged from coarse graining an underlying deterministic theory. One could not be witness to a more convincing demolition, than this one, of the `it from bit' hypothesis. 

It is noteworthy that Bohmian Mechanics, Spontaneous Localization and Trace Dynamics are three experimentally distinguishable `its' leading to the same `bit' [Figure 2]. While BM makes the same experimental predictions as standard quantum theory,  the predictions of CSL differ from those of quantum theory in the mesoscopic  domain. TD agrees with CSL at low energies but differs from it at energy scales approaching the Planck domain.

\begin{figure} [ht]
 {\includegraphics[width=\textwidth]{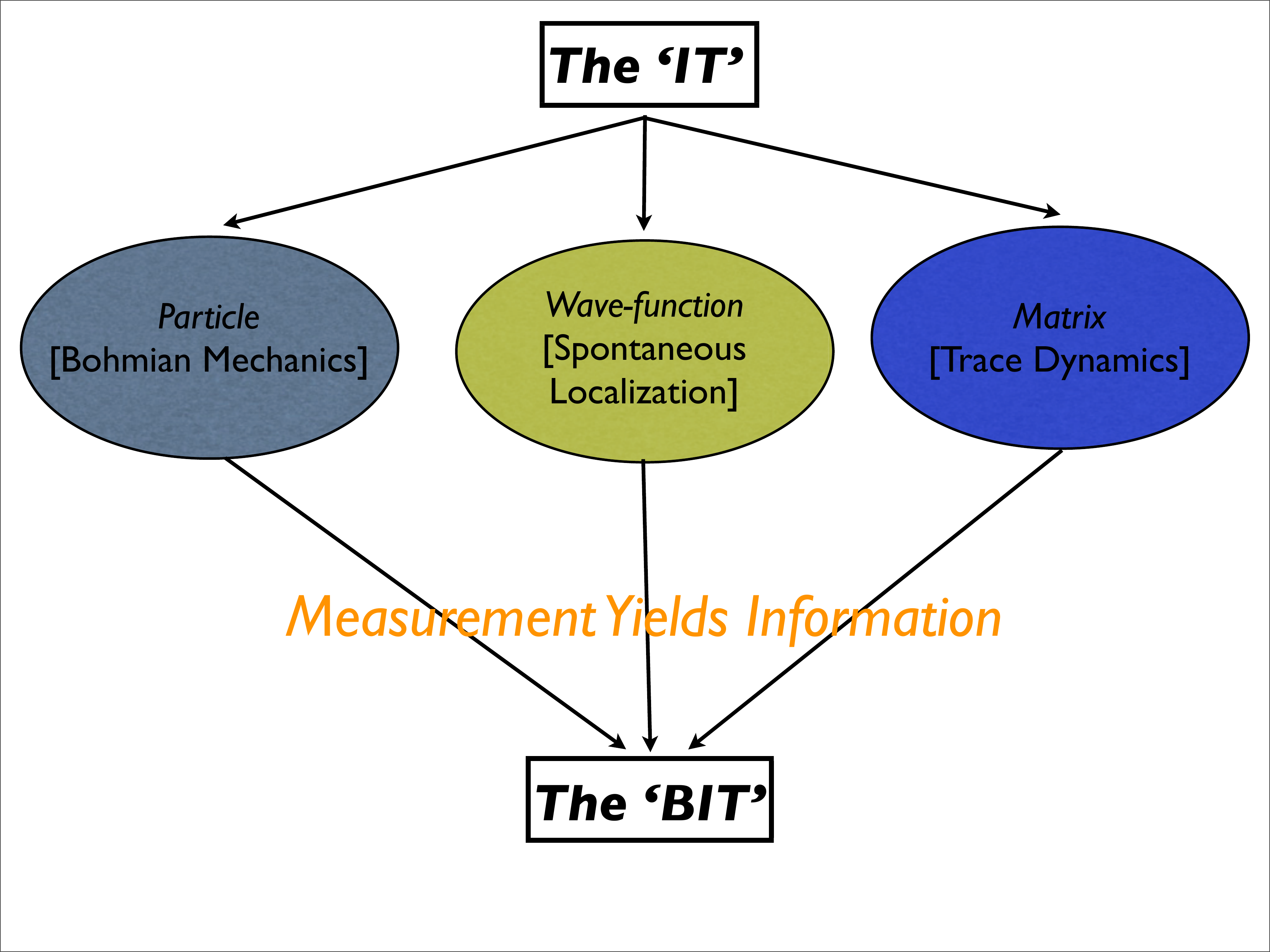}}
\caption{Bit from It: The Threefold Way: Bohmian Mechanics / Spontaneous Localization / Trace Dynamics}  
\end{figure}%

\section{Quantum theory without classical time}

The story would not be complete until one last barrier has been crossed. Quantum theory as we know it, depends on an external classical time. Such a time belongs to a classical spacetime geometry, which is produced by macroscopic bodies, which are themselves a limiting case of quantum theory. Thus, once again, via its dependence on time, quantum theory depends on its own limit. To improve matters, there ought to exist an equivalent formulation of quantum theory whch does not refer to an external classical time. [The same should in principle be required of CSL and Trace Dynamics as well.]

Imagine that one has at hand such a mathematical reformulation of quantum theory, by way of which one can describe a quantum system without referring to classical time. If one were to use this reformulation to describe measurement, would one again have to take recourse to irremovable probabilities? Would the `it from bit' return?

Fortunately not. There exists a Generalized Trace Dynamics [GTD] in which material degrees of freedom as well as spacetime are treated as matrices which together obey a classical dynamics. This is as primordial an `it' as an `it' can possibly get! Because every likelihood of irremovable probabilities has now been removed from the matter - spacetime arena. The GTD possesses a conserved charge akin to Trace Dynamics. The construction of an equilibrium statistical thermodynamics of GTD gives rise to a generalized quantum dynamics which does not refer to a classical time \cite{Lochan-Singh:2011, Lochan:2012}.  The consideration of statistical [Brownian] fluctuations around equilibrium allows for the emergence of classical spacetime and classical matter degrees of freedom, as well as the Born probability rule for different random outcomes for spacetime-matter configurations \cite{Singh:2012}.  Coarse graining is again at the root of probabilities, and `bit from it' now applies not only to matter quantum degrees of freedom, but also to quantized spacetime. [For a different view on the status of gravity in Trace Dynamics see  the recent work of Adler \cite{AdlerGravity}].

The greatest challenge to all the above four routes is that they are all non-relativistic. We do not till today have a relativistic theory of dynamical collapse, or a relativistic Bohmian mechanics.  Does this point to a fundamental limitation in the reach of these reformulations, and will the bit eventually reign over it? We do not know.  [It should be emphasized though that Trace Dynamics is fundamentally relativistically invariant; however the considerations which lead from the underlying theory to dynamical wave function collapse are at a non-relativistic level].

But we believe that "it from bit" is not a real option. "Bit" always refers to a pre-existing "it" [Figure 2]. This is the meaning of "bit". All confusion comes from inverting the order of "bit" and "it". When one starts the right way with the "it", then all problems evaporate.

This work is supported by a grant from the John Templeton Foundation.


\newpage

\centerline{\bf REFERENCES}

\bibliography{biblioqmts3}

\end{document}